\documentclass[prl,twocolumn,superscriptaddress,showpacs]{revtex4}
\usepackage{graphicx}
\newcommand{\beq}{\begin{equation}}
\newcommand{\eeq}{\end{equation}} 
\newcommand{\beqa}{\begin{eqnarray}}
\newcommand{\eeqa}{\end{eqnarray}}
\newcommand{\ba}{\begin{array}}
\newcommand{\ea}{\end{array}}
\begin{document}
\bibliographystyle{prbrev}

\title{Superfluid behavior of quasi-1D p-H$_2$ inside carbon nanotube}

\author{Maurizio Rossi}
\affiliation{Scuola Normale Superiore, 
			 Piazza dei Cavalieri 7, I-56126 Pisa, Italy}
\affiliation{International Center for Theoretical Physics (ICTP), 
			 Strada Costiera 11, I-34154 Trieste, Italy}

\author{Francesco Ancilotto}
\affiliation{Dipartimento di Fisica e Astronomia ``Galileo Galilei''
			 and CNISM, 
			 Universit\`a di Padova, via Marzolo 8, 35122 Padova, Italy}
\affiliation{CNR-IOM Democritos, 
			 via Bonomea, 265 - 34136 Trieste, Italy}

\begin{abstract}

We perform ab-initio Quantum Monte Carlo simulations of para-hydrogen 
(pH$_2$) at $T=0$ K confined in carbon nanotubes (CNT) of different radii.
The radial density profiles show a strong layering of the pH$_2$ molecules
which grow, with increasing number of molecules, in solid concentric 
cylindrical shells and eventually a central column.
The central column can be considered an effective one-dimensional (1D)
fluid whose properties are well captured by the Tomonaga-Luttinger liquid
theory.
The Luttinger parameter is explicitly computed and interestingly it shows 
a non-monotonic behavior with the linear density similar to what found for
pure 1D $^3$He.
Remarkably, for the central column in a (10,10) CNT, we found an ample
linear density range in which the Luttinger liquid is (i) superfluid and
(ii) stable against a weak disordered external potential, as the one 
expected inside realistic pores.
This superfluid behavior could be experimentally revealed in bundles of 
carbon nanotubes, where deviations from classical inertial values 
associated with superfluid density could be measured 
by using quartz crystal microbalance techniques.
In summary, our results suggest that pH$_2$ within carbon nanopores could 
be a practical and stable realization of the long sought-after, elusive 
superfluid phase of parahydrogen.

\pacs{67.25.D-, 67.25.dr}
\end{abstract}
\date{\today}
\maketitle

Superfluid para-hydrogen (pH$_2$) represents one of the most elusive 
phases in Nature.
As liquid helium, pH$_2$ is a natural candidate for displaying 
superfluidity by virtue of its light mass.
Contrarily to helium however, it does not remains liquid down to zero
temperature as a consequence of the stronger attractive interaction, 
which is about four times larger than the He-He one, and undergoes a 
crystallization transition around 14 K \cite{Cla06} (at saturated vapor 
pressure), a temperature much higher than the calculated superfluid 
bulk transition temperature $T\sim 1.1\,K$ \cite{Ape99}.
To create superfluid pH$_2$ it is therefore necessary to bring the liquid 
below its saturated vapor pressure curve. 
Attempts to produce a bulk superfluid pH$_2$ sample by supercooling the 
normal liquid below the triple point have been unsuccessful so far 
\cite{Mar83}. 

As a possible way to stabilize the liquid phase of pH$_2$ at low 
temperatures several authors have considered restricted geometries to 
reduce the effective attraction between molecules, and thus the 
zero-pressure density.
For example, the lowering of the melting point compared to the bulk liquid 
is a well-known and rather general phenomenon in clusters \cite{Alo05}, 
and a widely explored route in the search for pH$_2$ superfluidity is 
indeed based upon the realization and study of ultra-small pH$_2$ clusters.
The associated reduction of scale suggests that pH$_2$ clusters could 
display superfluidity. 
This expectation is based on the fact that the smaller number of neighbors 
and surface effects in small clusters may hinder solidification and promote
a liquid-like phase at low temperature\cite{Vil08}. 
The first prediction of superfluidity in pH$_2$ clusters made of few 
molecules was reported in 1991 based on quantum Monte Carlo (QMC) 
simulations \cite{Sin91}.
The first experimental signature of superfluidity in pH$_2$ was in fact 
inferred from the un-hindered rotation of a chromophore molecule in a 
cluster made of $N=15$ pH$_2$ molecules embedded in a larger $^4$He 
nanodroplet \cite{Gre00}.
Even if larger droplets of pH$_2$ are found to remain liquid at low 
temperature \cite{Vil08}, experiments of non-classical rotation seem to
locate the maximum size for a superfluid cluster at $N=17$ \cite{Roy10}.
Despite the great effort devoted to such systems \cite{Roy14}, any attempt 
of a direct observation of a stable superfluid phase of pH$_2$ has so far 
failed.

Another possibility put forward in theoretical calculations is to exploit 
disorder for suppressing crystallization and promote a superfluid response.
However, even in the most favorable scenario, disorder gives rise to 
a glassy phase which is predicted to be superfluid in a metastable regime 
\cite{boronat} but not at equilibrium \cite{Tur08}.

Taking advantage of the understanding gained for $^4$He systems, 
geometrical confinement has been considered too as a possible route to
stabilize a bulk superfluid phase for pH$_2$.
In fact, as inferred from extensive investigations for $^4$He in porous 
media such as Vycor \cite{Zas99}, zeolites \cite{Tod07}, and aerogel \cite{Yoo98},
as well as in superfluid films \cite{Nye98}, quantum fluids in constrained 
geometries behaves differently than in the bulk.
Equivalent indications of possible superfluid behavior for pH$_2$ in 
nanoconfined systems are scarce and often contradictory.
A possible superfluid phase inside a (5,5) carbon nanotube 
was predicted by studying the equation of state of pure one dimensional 
(1D) pH$_2$ at $T=0$ K with Diffusion Monte Carlo \cite{Gor00}.
However, recent Path Integral Monte Carlo (PIMC) calculations seem to 
contradict this claim, by showing that the 1D pH$_2$ equilibrium phase 
is a crystal \cite{Bon13}, as it also is in 2D pH$_2$ \cite{Bon04}.

So far the only reported enhancement of superfluid response was obtained
within PIMC simulations for pH$_2$ confined inside nano-cavities \cite{Omi14}.
The confining medium discussed in this paper is however not realistic,
being composed of spherical nano-sized cavities coated with alkali metal 
thick films in order to reduce the adsorption properties of the cavity 
walls, which seems hardly feasible at the present time.

We follow here a different approach by addressing a more realistic system 
made of pH$_2$ molecules in a confining system that is routinely provided 
by existing nanotechnologies, i.e. armchair carbon nanotubes (CNT) of 
different radii.
Although strictly 1D geometry precludes superfluid behavior, wider tubes 
where pH$_2$ forms a quasi-1D system coexisting with solid-like concentric
cylindrical shells could provide the ideal environment where strong evidence 
of the elusive superfluidity of para-hydrogen could be collected, as shown 
in the present work.

Our calculations are based on exact zero temperature Path Integral Ground 
State (PIGS) Monte Carlo method \cite{Sar00,Ros09}.
Because PIGS is a well-established computational methodology we shall 
not review it here.
We recall only that the most relevant feature is that it provides 
unbiased estimates of the $T=0$ K ground state properties directly by 
the microscopic Hamiltonian, by projecting in imaginary time a trial wave 
function.
The quality of the trial wave function has the sole role to fix the length
of the total imaginary time projection.
Here we have considered a shadow wave function (SWF) \cite{Vit88}, which
has provided an optimal trial wave function for bulk \cite{Gal03}, confined 
\cite{Ros12}, overpressurized \cite{Ros12b} and dimensionally reduced 
\cite{Vit08} $^4$He systems, whose parameters have been optimized to 
describe pH$_2$ \cite{Ope04}.
All the approximations involved in the PIGS method, i.e. the choice of 
the total imaginary time $\tau$, of the imaginary time step $\delta\tau$
and the approximation for the short imaginary time propagator, are so 
well controlled that the resulting systematic errors can be reduced within 
the unavoidable Monte Carlo statistical error making of PIGS an {\it exact} 
zero-temperature method \cite{Sar00,Ros09}.

In our calculations we consider $N$ pH$_2$ molecules, described as 
point-like particle with zero spin adsorbed within CNT of different radii, 
described by the following Hamiltonian
\begin{equation}
 \label{H}
 \hat H = -\lambda \sum_i \nabla_i^2 
          + \sum_{i<j}v(|\vec r_i - \vec r_j|)
          + \sum_i V(\vec r_i),
\end{equation}  
where $\vec r_i$ are the positions of the pH$_2$ molecules, 
$\lambda = \hbar^2/2m = 12.031$ K\AA$^2$, $v$ describes the interaction
between a pair of molecules and $V$ describes the interaction of a molecule
with the CNT.
We assume periodic boundary conditions along the tube axis.
As for $v$, which is considered spherically symmetric, we use the well-known
Silvera-Goldman potential (SG) \cite{Sil78}.
We specifically consider three different armchair CNTs: the $(10,10)$ CNT 
with radius $R=6.80$ \AA, the $(12,12)$ CNT with radius $R=8.19$ \AA\ 
and the $(15,15)$ CNT with radius $R=10.17$ \AA.
To model the H$_2$-carbon interaction we used a pair potential fitted 
to high level ab initio results on the interaction between H$_2$ and 
graphite \cite{Sun07}.
This procedure provides a corrugated potential $V$, but implicitly neglects 
the effects of curvature, which however are found to have very little 
consequences for the considered CNTs \cite{Suppl}.
By using the fourth order pair-Suzuki approximation \cite{Ros09} for the 
short imaginary time propagator we observe convergence of ground state
estimates with a projection time $\tau = 0.250$ K~$^{-1}$ using a time 
step $\delta\tau = 1/640$ K~$^{-1}$. 

\begin{figure}[t]
 \includegraphics[width=1\linewidth,clip=true]{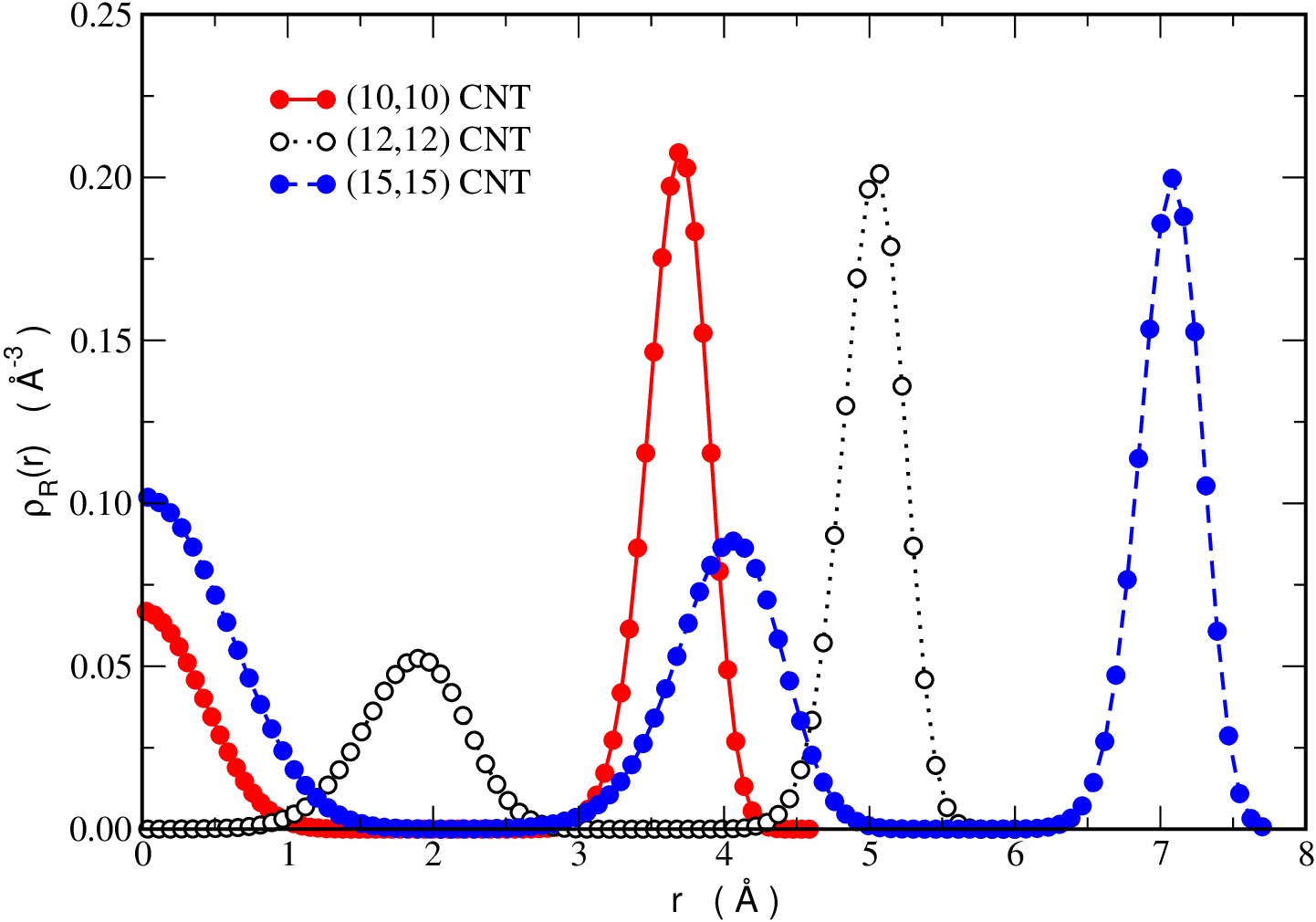}
 \caption{\label{fig1} (Color online) Typical radial density profiles 
 					   $\rho_R(r)$ for pH$_2$ inside the considered CNTs. 
   					   Statistical errors are smaller than the used symbols.}
\end{figure}

We made a number of simulations with varying number $N$ of pH$_2$ molecules
from 38 up to a maximum of 432, adsorbed within the three different CNTs
of increasing length $14.77 < L < 59.00$ \AA.
Similarly to the case of $^4$He adsorbed in nanotubes 
\cite{Ros05,Ros07,Mae11,Mae12,Kul13,Pol14}, the pH$_2$ radial density profiles 
$\rho_R(r)$, reported in Fig.~\ref{fig1}, show a marked layered structure: 
the pH$_2$ molecules form a cylindrical shell adsorbed on the inner tube 
wall plus a central column in the (10,10) CNT, two concentric cylindrical 
shells in the (12,12) CNT, and two concentric cylindrical shells plus a 
central column in the (15,15) CNT.
The promotion to the second (lower density) layer (or to the central column 
for the (10,10) CNT) occurs at an areal density value $\theta_P = 0.108\,\AA^{-2}$, 
which is about 15\% higher than the promotion coverage to the second layer 
found for pH$_2$ adsorbed on graphite \cite{Bru97}.
For all the three CNT, the adsorbed layer adjacent to the tube inner surface 
turns out to be a crystalline two dimensional triangular solid wrapped 
to form a cylinder, whose structure is incommensurate with the underlying 
carbon lattice.
The intermediate shells (in the (12,12) and (15,15) tubes) are also two 
dimensional solid wrapped on a cylindrical surface for all the considered 
values of $N$, with no evidence whatsoever of a liquid-like behavior.

Within the PIGS method it is possible to obtain a direct estimate of the
superfluid fraction $\rho_s$ from the imaginary time diffusion of the 
center of mass of the system \cite{Zha95,Nav12}.
For the (10,10) and (12,12) CNTs, we found a sizable superfluid response,
comparable with the (debated \cite{Bon16}) one calculated for pH$_2$ 
embedded in a 2D Na crystal \cite{Caz13}.
However, this large $\rho_s$ has to be ascribed to the presence of defects
in the crystalline structure of the adsorbed layer due to the mismatch 
between the pH$_2$ lattice and the underlying carbon structure, i.e. to 
the actual length $L$ of the simulated CNT.
The effective ability of such defects to sustain a detectable superfluid 
flow \cite{Ros10} or rather their pinning at the structural defects in 
real CNTs is beyond the scope of this paper.

Our systems provide however a better candidate for superfluidity: the 
central column in (10,10) and (15,15) CNTs, that behaves as a quasi-1D 
superfluid whose properties are well captured by the Tomonaga-Luttinger 
liquid theory (TLL) \cite{Tom50,Lut63,Hal81}.
The description of a confined quantum fluid by means of the TLL has been 
successfully applied to $^4$He in nanopores \cite{Mae11,Mae12}.

TLL is a phenomenological theory than captures the low-energy properties 
of a wide class quantum 1D systems with short range interaction \cite{Tom50,Gia03}
in terms of two bosonic fields, $\phi(x)$ and $\theta(x)$ representing 
respectively the density and the phase fluctuations of the particle field 
operator $\psi(x) = \sqrt{ \rho + \partial_x\phi(x) } \, e^{i \theta(x)}$ 
($\rho$ being the average density) via the low-energy effective Hamiltonian 
\begin{equation}
 \label{HLL}
  H_{LL} = \frac{\hbar}{2\pi} \int dx \, \left( c K_L \partial_x\theta(x)^2 
         + \frac{c}{K_L} \, \partial_x\phi(x)^2 \right) \quad.
\end{equation}
The parameter $K_L$ (known as Luttinger parameter \cite{nota1}) and the 
velocity $c$ are generally independent quantities fixed by the microscopic 
details of the system.
Such Hamiltonian is exactly solvable, thus the knowledge of $c$ and $K_L$ 
is enough to characterize the correlation functions and the thermodynamic 
properties of the system.
For Galilean-invariant systems, as the ones we are going to consider here,
$c=\hbar\pi\rho / m K_L$ \cite{Hal81}, thus the only parameter to be 
determined is $K_L$.
Here we determine the Luttinger paramenter via QMC simulations, that have 
largely been proven to ba efficient in estimating $K_L$ \cite{Bon13,Mae10,Ast14,Ber14}.

$K_L$ governs the decay of correlations function and can be used to draw 
a well defined definition of (quasi) crystal and (quasi) superfluid.
For $K_L<1/2$ the static structure factor develops Bragg peaks at reciprocal 
lattice vectors, which is the signature of a (quasi) crystalline solid.
For $K_L>1/2$ no (quasi) diagonal-long range order is present, but the
system displays a (quasi) off-diagonal long-range order.
Thus, even if no true long-range order can exist in 1D for a system of 
particles with short range interaction \cite{Mer66}, there can be a 
phase, known as Luttinger liquid (LL), featuring power-law decaying 
correlations \cite{Hal81}, that is superfluid in the sense that it displays 
a quasi-off diagonal long range order \cite{Egg11,Caz11}.
Such a superfluidity manifests different degree of robustness against
disorder or external potentials.
Specifically, if $K_L>3/2$ the superfluid is insensitive to a weak external
disordered potential \cite{Gia87}, while for a periodic external potential
commensurate with the density with filling fraction $1/p$, the transition
is located at $K_L=2/p^2$ \cite{Caz11}.

\begin{figure}[t]
 \includegraphics[width=1\linewidth,clip=true]{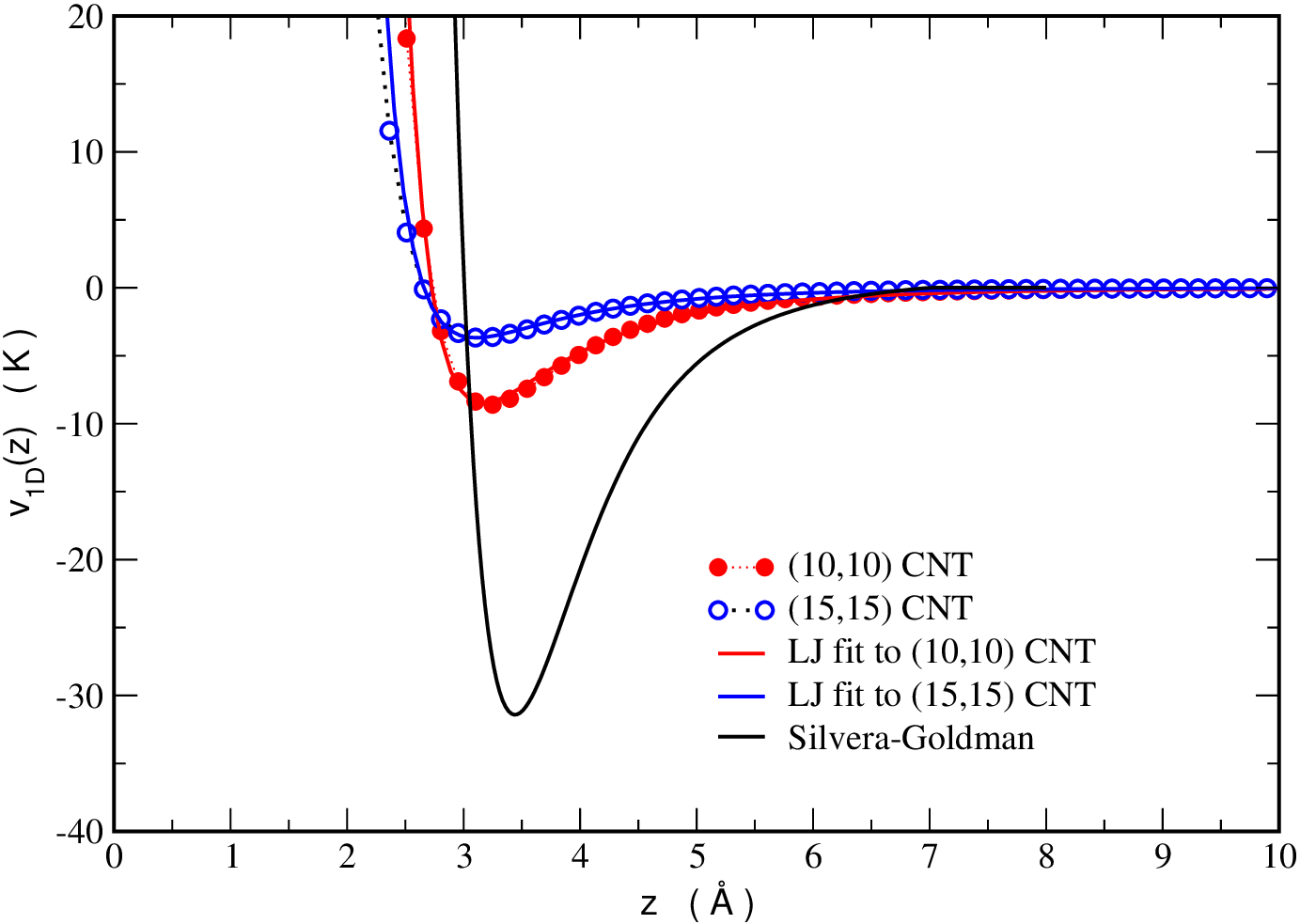}
 \caption{\label{fig2} (Color online) Effective one-dimensional interaction
potential $v_{\rm 1D}$, obtained by Eq.~(\ref{v1D}), 
for pH$_2$ molecules in the central column inside 
(10,10) and (15,15) CNT. 
The effective potentials are compared with the 
Silvera-Goldman potential for the pH$_2$-p$H_2$ 
interaction.
Continuous lines represent a fit of $v_{\rm 1D}$ 
with a Lennard-Jones (LJ) like potential.}
\end{figure}

It was found that for narrow pores $^4$He obeys the TLL theory with a 
small Luttinger parameter corresponding to solid-like character 
of the adsorbed phase \cite{Mae11,Mae12}. 
On the other hand, for wider pores, the central region appears to behave 
like a LL but with a larger $K_L$ indicating that the system is dominated 
by superfluid fluctuations, as indeed expected for superfluid $^4$He.
We will show here that a similar behavior occurs for p$H_2$ in CNTs.
Since the exchanges of molecules with the surrounding shells are null, 
the central column of pH$_2$ in (10,10) and (15,15) CNT can be considered 
an effective one-dimensional system that can be well described via the 
TLL \cite{Mae12}.
The surrounding shells have the crucial role to screen (reduce) the bare 
pH$_2$-pH$_2$ interaction, and the molecules inside the central column can 
be depicted as pure 1D particles (with the same mass of the initial ones) 
interacting via the effective potential \cite{Mae12}
\begin{equation}
 \label{v1D}
  v_{\rm 1D}(z) = \frac{1}{\rho_L^2} \int d^2r \int d^2 r' v(\vec r - \vec r')
  \rho_R(r)\rho_R(r')
\end{equation}
where $\vec r = (r,\varphi,z)$ is a vector in cylindrical coordinates and
\begin{equation}
 \label{rhoL}
 \rho_L = \frac{N}{L}=2\pi \int dr\ r\rho_R(r)
\end{equation}
is the linear density.
The resulting effective potential is almost insensitive to the actual 
pore length and to the density of the central column itself.
The obtained $v_{\rm 1}$ for the (10,10) and (15,15) CNT are shown in
Fig.~\ref{fig2}, where they are also compared to the SG pH$_2$-pH$_2$
bare interaction potential.
Notably, they can be fairly fitted by using a Lennard-Jones like formula 
\cite{nota3}.
As already observed for $^4$He \cite{Mae12}, the effect of the surrounding
shells is that of reducing the potential well depth and of shifting the
minimum to smaller separations.
This can change dramatically the low-density behavior of the effective 1D
system when compared to the strictly 1D pH$_2$.
In fact, for example, pure 1D pH$_2$ is expected to display a spinodal 
decomposition at densities below 0.209 \AA$^{-1}$ \cite{Bon13}, while we
have been able to simulate such effective pure 1D system down to 
$\rho=0.02$ \AA$^{-1}$ without any signature of spinodal decomposition.
For this pure 1D system, we have simulated $N=50$ particles in order to 
minimize the finite size effects \cite{Bon13,Ber14}, taking 
$\tau = 2.50$ K~$^{-1}$ and $\delta\tau = 1/320$ K~$^{-1}$ to guarantee
convergence to the ground state. 

\begin{figure}[t]
 \includegraphics[width=1\linewidth,clip=true]{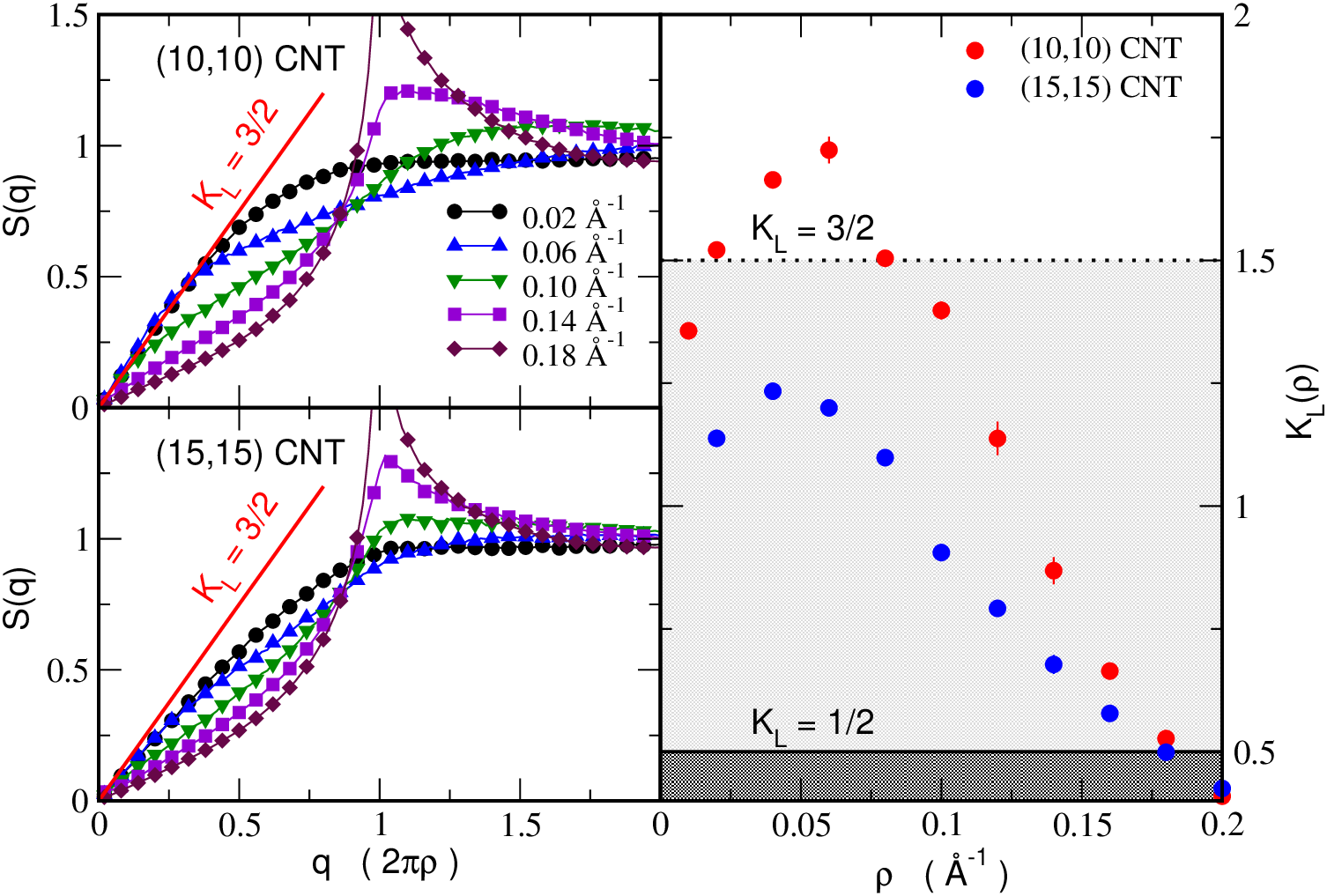}
 \caption{\label{fig3} (Color online) Left: static structure factors S(q) 
                       for both the effective 1D systems realized by the 
                       central column of pH$_2$ inside the (10,10) and the 
                       (15,15) CNT for different linear densities $\rho$.
                       The momenta $q$ are given in units of $2\pi\rho$.
                       Errorbars are smaller than the used symbols.
                       The straight red line marks the threshold $K_L=3/2$.
                       Right: Luttinger parameter $K_L$ as obtained by the
                       low momenta behavior of the static structure factor.
                       Both the relevant thresholds $K_L=1/2$ (which discerns
                       from a (quasi) crystal and a (quasi) superfluid) and 
                       $K_L=3/2$ (which marks the stability of the superfluid
                       against a weak disordered external potential) are 
                       also shown.}
\end{figure}

The Luttinger parameter $K_L$ can be extracted by the low momenta behavior
of the static structure factor $S(k)$ \cite{Ast14,Ber14}
\begin{equation}
 \label{sdk}
 S(k)=\frac{K_L k}{2\pi\rho},\ \ \ \ k\to0.
\end{equation}
Some example of the calculated $S(k)$ for the effective 1D systems 
realized by the central column of pH$_2$ inside the (10,10) and the (
15,15) CNT are reported in Fig.~\ref{fig3} for different values of the 
linear density $\rho$.
The linear behavior at low momenta is evident and the extracted $K_L$ are 
reported in the right panel of the same figure.
The dependence of $K_L$ from the density is non-monotonic and resembles
the one for $^3$He \cite{Ast14}.

The key result of the present work is however that for the central column 
of pH$_2$ inside a (10,10) CNT there is an ample density range 
($0.02 < \rho < 0.08$ \AA$^{-1}$) where $K_L>3/2$, meaning that the LL is 
both superfluid and stable against the presence of a weak disordered 
external potential.
The stability of such a 1D superfluid is crucial because pH$_2$ inside
the central column is expected to indeed experience an external potential.
For the (10,10) CNT, the potential provided by the CNT itself is practically 
flat on the pore axis, while the one provided by the surrounding solid 
layer is weak and disordered because of incommensurability effects \cite{nota4}.
No stable LL superfluid has been observed for the (15,15) CNT, where the
$K_L$ is always lower than $3/2$.
It is interesting to note that the central columns in both the CNT undergo
a crystal-superfluid transition at linear densities closed to the spinodal
decomposition of the pure pH$_2$.

A possible way to experimentally stabilize the superfluid phase of pH$_2$ 
is by confining it within aligned bundles of micron-sized parallel CNTs.
The predicted superfluid phase of pH2 could be observed by using current 
quartz microbalance techniques \cite{Suppl}, by measuring the frequency 
shifts in the shear modes of the microbalance parallel to the bundle axis.
The density range where a superfluid response should be expected could be 
easily reached by changing the pressure (chemical potential) of the pH$_2$ 
vapor surrounding the nanotube bundles, which determines the actual amount 
of fluid adsorbed inside the central columns.

While preparing this manuscript, we were made aware of a recent paper
where quasi-1D pH$_2$ in model nanopores with smooth walls was studied 
by using PIMC \cite{Omi15} with no sign of any LL-superfluid phase.
When their radius is such that the pore houses a single pH$_2$ column, 
$K_L$ is found to grows from the value 0.28 of the pure 1D pH$_2$ 
\cite{Bon13} to values close to $1/2$ but still in the (quasi) crystalline 
state.
Even the combination of cylindrical shell plus central column has been
explored in Ref.~\cite{Omi15} for a glass pore of radius $R = 5$ \AA, but 
with opposite results than ours.
We argue that the less attractive substrate and the small radius
with respect to the (10,10) CNT considered here, provide a tighter 
confinement for pH$_2$ that strongly localizes the molecules, resulting 
in a Luttinger parameter even lower than the one of the pure pH$_2$.

In conclusion we have studied, by using PIGS exact simulations and the 
TLL theory, pH$_2$ within geometric confinement provided by realistic CNT 
of different radii.
The results show the appearance, for the (10,10) and (15,15) CNTs, of a 
central column along the NT axis, that can be described within the TLL 
theory.
The molecules belonging to the inner column are screened by the solid 
pH$_2$ layers adsorbed on the surrounding CNT inner wall, resulting in 
pronounced quantum exchanges between molecules within the column, 
leading to a clear superfluid behavior in the (10,10) CNT.
Our QMC simulations do indeed confirm this scenario, suggesting that 
pH$_2$ within bundles of carbon nanotubes could be a practical 
realization of the elusive superfluid phase of parahydrogen.

We thank M. Boninsegni, G. Bertaina, D.E. Galli, G. Mistura and P.L. 
Silvestrelli for stimulating discussions.

\newpage

\section{SUPPLEMENTAL MATERIAL: Superfluid behavior of quasi-1D p-H$_2$ inside carbon nanotube} 

\bigskip
\bigskip
\bigskip

\section {pH$_2$-CNT interaction potential}

The effective pH$_2$-CNT interaction potential $V$ is constructed by 
summing all the pairwise pH$_2$-C contributions among the pH$_2$ molecule 
and the Carbon atoms of an armchair CNT tube of indices (n,n) and stored
on a grid for computational efficiency.
To prevent spurious boundary effects due to the truncation of the tube 
(and on the imposed periodic boundary conditions along the axis) we 
construct tubes longer than 100 \AA\ to perform the summation but retain 
only the central portion of length $L_0=14.77$ \AA\ (corresponding to 5
elemental cells of the Carbon hexagonal lattice).
Thus the simulated CNTs described in the paper will have lengths that 
are integer multiples of $L_0$.

\subsection{Curvature effects.}

In order to describe the pH$_2$-C interaction we have considered an 
exp-6-8-10 potential obtained by fitting high-quality {\it ab initio} 
results for the H$_2$-graphite interaction \cite{Sun07}.
Although our potential construction accounts for the cylindrical geometry
of the C atoms, it misses the curvature related $sp^2 \rightarrow sp^3$ 
hybridization of the Carbon bonds that may modify the pH$_2$-C interaction.
Curvature-dependent corrections to the coefficients of a Lenard-Jones 
potential describing the C-H$_2$ pairwise interaction have been proposed 
in the past to include effects due to the increasing $sp^2 \rightarrow sp^3$ 
hybridization in Carbon nanotubes as the CNT radius decreases\cite{Kos02}.
Such corrections are based on interpolation formulas between the two 
extreme limits of pH$_2$ interacting with a C atom in a graphite plane 
(pure $sp^2$ character) and in an ideal $sp^3$ environment. 
However, this model potential was later found inadequate to quantitatively 
describe the H$_2$-graphite interaction and thus a better, albeit curvature 
independent, C-H$_2$ pair potential has been proposed \cite{Sun07}, which
we are using here.
Corrections due to mixed hybridization effects mentioned above (which are 
important for CNTs with very small radii) are expected however to have 
small impact for the CNT investigated in our work. 
In fact, for the radii of interest here, the $sp^2$ component in the 
interaction parameter predicted by the interpolation formulas of 
Ref.~\cite{Kos02} would be dominant (being 86\% for the (10,10) tube and 
91\% for the (15,15) one). 
We should also add that among physisorbed system, Carbon materials
represent rather attractive substrates for pH$_2$ adsorption, and thus 
relatively small changes in the adsorption well depth/position of the 
pH$_2$-C potential are expected to affect only slightly the structure of 
the solid-like pH$_2$ layer adsorbed on the CNT inner wall, and have even 
smaller effects on the central column structure. 

In order to check that curvature effects are indeed negligible in the 
present study, we proceeded as in Ref.~\cite{Kos02} to compute the 
interaction between H$_2$ and an $sp^3$-coordinated Carbon atom with a 
dangling bond, and then use their interpolation formulas to find the 
correction to the pair potential well-depth used in our calculation 
(appropriate for a flat, $sp^2$ coordinated substrate). 
We performed state-of-the-art {\it ab initio} calculations (using the 
QUANTUM-ESPRESSO package\cite{QE}) to compute the interaction between a
H$_2$ molecule and a t-butyl radical (used in Ref.~\cite{Kos02} to model 
the environment of an unsaturated $sp^3$ carbon). 
The exchange-correlation functional used in our calculations explicitly 
includes the contribution of dispersion (Van der Waals) forces \cite{Lee10} 
(which were neglected in the calculations of Ref.~\cite{Kos02}), which 
are known to provide important corrections to the energetics of 
physisorbed systems like the one considered here.
We find that in order to reproduce the ab-initio results
described above for the H$_2$-C($sp^3$) interaction, 
the well depth of the exp-6-8-10 pair potential
should be $\epsilon = 6.2\,meV$, to be compared to the 
value $\epsilon = 5.3\,meV$ describing the pH$_2$-C($sp^2$) two-body interaction.
By using the above values in the interpolation formula that gives the curvature-corrected 
well depth parameter of the two-body interaction appropriate for a 
CNT with a given radius\cite{Kos02}, 
we find that in the case of the (10,10) tube, 
the pair-potential well depth should be decreased by 23\% with respect 
to the value (appropriate for the H$_2$-C($sp^2$) interaction)
used in our calculations. An even smaller change is found for the (15,15) tube.
In order to estimated the consequences of such a reduction, we compare 
the results of the QMC calculations for the radial density profile in the 
(10,10) CNT by summing (i) the pair potential without curvature corrections
$V$ (i.e. the one used for the computation in the main paper) and (ii) a 
rescaled one $\eta V$ with $\eta=0.9$, $0.8$ and $0.7$.
\begin{figure}[t]
 \includegraphics[width=1\linewidth,clip=true]{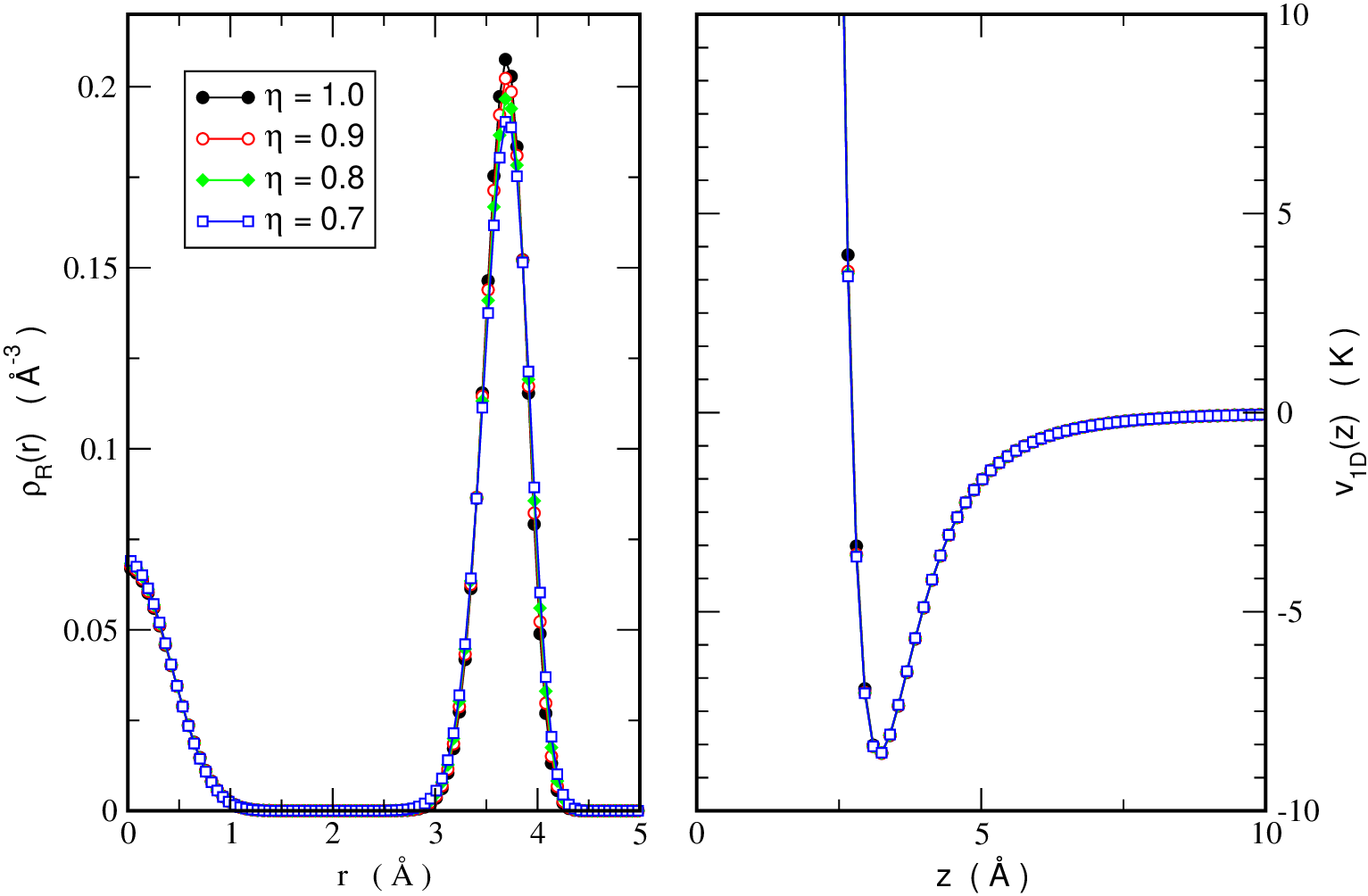}
 \caption{\label{figS1} (Color online) Left panel: radial density profiles 
 					   $\rho_R(r)$ for pH$_2$ inside a (10,10) CNTs with
 					   the potential $V$ rescaled by a factor $\eta$ as
 					   described in the text. 
   					   Statistical errors are smaller than the used symbols.
   					   Right panel: corresponding effective one-dimensional 
   					   interaction potential $v_{\rm 1D}$.}
\end{figure}
We find, as it appears in Fig.~\ref{figS1} where our results are 
reported, that the differences in the two cases (i.e. with and without 
curvature corrections) are indeed very small. 
The maximum of the density peak is reduced by a factor about 9\% for the
largest considered $\eta$, but even in this case the number of pH$_2$
molecules housed by the adsorbed layer is the same as for $\eta = 1.0$.
Since the curvature effects on the density profiles are small, the obtained
effective one-dimensional interaction potential $v_{\rm 1D}$ (Eq.(3) in the
main paper) is practically insensitive to such corrections, as can be 
inferred from the right panel of Fig.~\ref{figS1}.
This confirms that curvature effects are completely negligible for the
present study.

\section{Possible experimental setup to observe pH$_2$ superfluid behavior 
         in CNTs.}

Aligned structures of CNTs (of lengths up to hundreds microns) with 
controllable radii can be routinely grown on different substrates (see for 
instances Ref.~\cite{Che14}, and references therein). 
Micrometer-sized bundles of aligned parallel carbon nanotubes have been 
grown, in particular, on silicon or quartz surfaces \cite{uspat}, which 
can be up to 300 micron tall, with a bundle aspect ratio (height-to-width) 
of 5:1. 
This is a particularly favorable method in the present context, which 
allow in principle to grow the CNTs bundle directly over a quartz crystal 
microbalance (QCM) surface. 
Once filled with hydrogen (whose amount can be controlled through the 
environmental pressure), frequency shifts in the shear modes of the 
microbalance parallel to the bundle axis should allow detection of mass 
decoupling due to the appearance (e.g. by lowering the temperature) of 
a superfluid fraction within the central column of pH$_2$ in each CNT. 
The latter corresponds to about 3\% of the total amount of pH$_2$ adsorbed 
in a (10,10) CNT.
For a QCM driven at a fundamental frequency of 5 MHz, adsorption of pH$_2$ 
in the nanotubes will determine a frequency shift of 30 Hz if the total 
open volume of the nanotubes amounts to, say, $10^{-7}\,cc$ (which is 
well below the theoretical values accessible in the microscopic bundles 
described above). 
If about 3\% of the total pH$_2$ mass decouples from the QCM oscillations 
in the superfluid phase, this will determine an increase in the resonance 
frequency of about 1 Hz, which is well above the standard resolution of 
0.1 Hz (or less) typical of a QCM setup (see, for instance, Ref.~\cite{Bru01}), 
thus allowing, in principle, to detect the  normal-to-superfluid transition.


\begin{thebibliography}{99}
\bibitem{nota1} Two predominant definitions of the Luttinger parameter $K_L$ 
				exist in the literature. The one used here is the 
				inverse of the one employed in Ref.~\cite{Bon13,Mae11,Mae12,Kul13}.
\bibitem{nota3} We have fit our $v_{\rm 1D}$ with the formula
				$v_{LJ}(z)=4\varepsilon\left[\left(\frac{\sigma}{z}\right)^m 
                  - \left(\frac{\sigma}{z}\right)^{m/2}\right]$
				obtaining $\varepsilon = 8.485$ K, $\sigma = 2.726$ \AA\ and 
				$m = 9.230$ for the (10,10) CNT and $\varepsilon = 3.683$ K, 
				$\sigma = 2.658$ \AA\ and $m = 8.957$ for the (15,15) one.
\bibitem{nota4} The external potential experienced by the pH$_2$ molecules
				in the central column is the sum of the potential due to
				the CNT itself and the one due to the adsorbed layer.
				The CNT produces a periodic potential whose maximum excursion 
				in energy (i.e. the energy difference between top and
				hollow site adsorption energies) are of the order of 0.03 K, 
				i.e. $10^{-3}$ times the SG well and $4\times 10^{-3}$ times
				the $v_{\rm 1D}$ well.
				The adsorbed layer produces a potential which is disordered
				due to the presence of defects induced by incommensurability,
				whose maximum excursion in energy in 0.3 K, i.e. $10^{-2}$ 
				times the SG well and $4\times 10^{-2}$ times the $v_{\rm 1D}$ 
				well.
\end{thebibliography}
\end{document}